\documentstyle[sprocl_pre,a4,12pt,epsfig]{article}

\def\be{\begin{equation}}
\def\ee{\end{equation}}
\def\bea{\begin{eqnarray}}
\def\eea{\end{eqnarray}}

\begin{document}

\title{LAST MINUTE FROM ALADIN:Temperature measurements 
in Au+Au reactions at relativistic energies
\footnote{\normalsize To appear in the proceedings of the 1st Catania
    Relativistic Ion Studies: Critical Phenomena and Collective Observables,
    Acicastello, May 27-31, 1996.}
}

\author{ G. IMME', V. MADDALENA, C. NOCIFORO, {\underline{G. RACITI}},
 G. RICCOBENE, F.P. ROMANO, A. SAIJA, C. SFIENTI, G. VERDE }

\address{Dipartimento di  Fisica -
Universit\'a di Catania and I.N.F.N.- Laboratorio Nazionale del Sud  and
Sezione di Catania\\ 
I-95100 Catania, 
Italy}

\author{ M. BEGEMANN-BLAICH, S. FRITZ, 
C. GRO\ss, U. KLEINEVO\ss, U. LYNEN,
M. MAHI,
W.F.J. M\"ULLER, B. OCKER, T. ODEH, M. SCHNITTKER, C. 
SCHWARZ, V. SERFLING,
W. TRAUTMANN , A. W\"ORNER, H. XI}

\address{Gesellschaft f\"ur Schwerionenforschung mbH, 
Planckstrasse,1\\
D-64291 Darmstadt, Germany}

\author{ T.~M\"OHLENKAMP, W. SEIDEL }

\address{Forschungszentrum Rossendorf ,\\
D-01314 Dresden, Germany}

\author{ W.D.KUNZE,A. SCH\"UTTAUF }

\address {Institut f\"ur Kernphysik, Universit\"at Frankfurt,\\
D60486 Frankfurt,Germany}
 
\author{ J. POCHODZALLA }

\address{Max Planck Institut Heidelberg ,Saupfercheckweg 
1\\
D-69177 Heidelberg, Germany}

\author{ G.J. KUNDE, S. GAFF }

\address{National Superconducting Cyclotron Laboratory 
Michigan State 
University ,\\
East Lancing,MI 48824, USA}

\author{ R. BASSINI, I. IORI, A. MORONI, F. PETRUZZELLI}

\address{I.N.F.N. Sezione di Milano and Dipartimento di 
Fisica, Universit\'a di
Milano,\\
I-20133 Milano, Italy}

\author{ A. TRZCINSKI, B. ZWIEGLINSKI }

\address{Soltan Institue for Nuclear Studies ,\\
00-681 Warsaw,Hoza69, Poland}

\maketitle\abstracts{We report on temperature 
measurements of nuclear systems
formed in the Au+Au collisions at incident energies of 
50, 100, 150, 200 and 
1000 A MeV.The target spectator matter was studied at 
the highest energy and
the interacting zone (participants) at the lower 
ones.The temperature deduced
from the isotope ratios was compared with the one 
deduced via the excited 
states population. An unexpected disagreement was found 
between the two measurements.}
  
\section{Introduction}
Previous studies on the systems produced in Au + Au 
collisions at
bombarding energies in the range 100- 1000 A MeV had 
shown that the decay of 
excited spectators proceeds independently of the 
entrance channel dynamics.
\cite{HUB91,KRE,OGI,SCHU}
In addition, strong indications of collective flow 
\cite{KUN95,HSI94} ,that 
characterizes the multifragmentation of the participant 
region, have not been found
in the decay of the spectator matter.
Therefore, projectile and target spectators, which are 
produced over a wide 
range of excitation  energy and mass in these reactions, 
are well suited for 
the investigation of thermally driven phase transition in highly excited 
nuclear systems.
In particular, the correlation between measured 
excitation energies of the 
decaying projectile spectator and emission temperature 
of the produced fragments
,measured via the double ratios of isotope 
yields\cite{ALB}, for Au+Au at 600 
A MeV of incident energy, allowed us to deduce a caloric 
curve of nuclei
\cite{POC95}.
The observed behaviour of the caloric curve is
reminiscent of a first-order phase transition in 
macroscopic systems.
A region of constant temperature of T $\approx$ 5 MeV 
separates the
'liquid' and 'vapour' regimes where the temperature 
rises with
increasing excitation energy. These data have been 
subjected to a widespread 
discussion which addresses both, methodical aspects and 
questions of
interpretation \cite{MOR95,NAT95,TSA95}.Therefore, in 
order to 
permit a crosscomparison of two different 'thermometers' 
which is 
mandatory for a quantitative interpretation of the 
results,
we studied collisions of Au + Au at two regimes of 
bombarding
energies: 1 GeV per nucleon to investigate the 
spectator decay, and mid-central
 collisions at 50, 100, 150, and 200 MeV per nucleon to 
study
the decay of the participant region which is known to 
exhibit a considerable 
collective flow.In all these studies we measured 
simultaneously both isotope 
yields and particle- particle correlations from the 
decay of unstable states 
in the emitted fragments. The latter method has been widely
 used at present
to achieve temperature 
measurements\cite{POC87,STL88,NAY92,XI93,KUN93,SCH93}.
One of the goals of the present experiment was to 
compare the temperature 
derived applying these two 
methods to the same class of events of the decay of the 
same nuclear system.

\begin{figure}[tb]
 \centerline{\epsfig{file=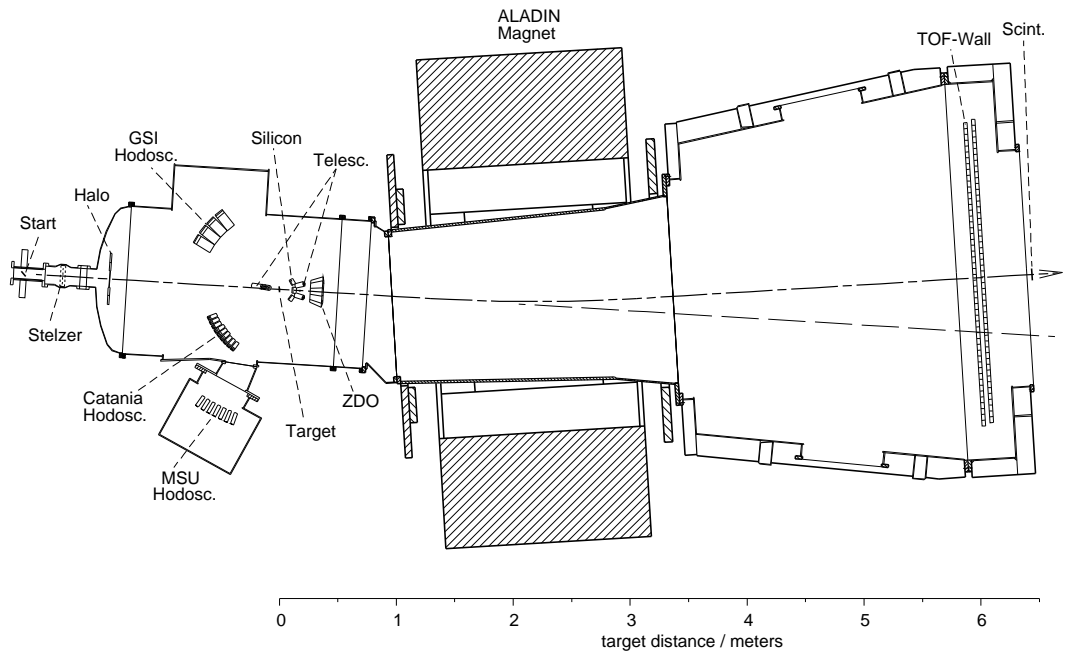,height=\textwidth,
bbllx=100,bblly=100,bburx=520,bbury=760,angle=-90}}
  \caption[]{
    Schematic view of the experimental setup at the ALADIN spectrometer
    for the measurement with Au beams of 1 GeV per nucleon.
  }
  \label{s117-setup}
\end{figure}

\section{Experimental Apparatus}
The experiment was performed at the SIS accelerator of 
the GSI, using the
ALADiN spectrometer facility. A schematic
view of the experimental setup is shown in Fig. 
\ref{s117-setup}.
Three hodoscopes, from GSI (64 elements), 
MSU (56 elements), 
and Catania(96 elements), consisting 
of Si-CsI(Tl) telescopes(Si 3x3 cm $^2$ 
300$\mu$m thick,CsI 
scintillators
3x3 cm $^2$ 6 cm long with  photodiode readout), were 
set up at distances
between 0.6 m and 1.1 m from the target. Their angular 
resolution and
granularity were optimized in order to permit the 
identification of
excited  particle-unstable resonances in light fragments 
from the
correlated detection of their decay products. For the 
measurements at
lower energies, the hodoscopes were placed  in the 
forward
hemisphere in order to cover the angular region close to 
90$^0$ in the
center-of-mass system. At 1 GeV per nucleon( Fig.1), the 
hodoscopes were moved
to the backward hemisphere in order to select mainly 
target spectator 
products. A set of seven
telescopes, each consisting  of three Si detectors with 
thicknesses
50~$\mu$m, 300~$\mu$m, and 1000~$\mu$m, followed by a  
4~cm long CsI
detector, was placed at selected angles. They served to
simultaneously measure the yields of isotopically 
resolved light
fragments. 

At 1 GeV per nucleon the impact-parameter selection was 
achieved by
measuring the quantity $Z_{bound}$ through the forward 
angles time-of-flight 
wall of the ALADiN spectrometer.
$Z_{bound}$ is defined as the sum of the atomic numbers 
$Z$ of all
fragments with $Z \ge$ 2.\cite{HUB91}.
Since the hodoscopes are set up at backward angles for
the detection of fragments emitted by the target
spectator, the method  used for impact parameter 
selection
exploits the symmetry of the collision system and avoids 
autocorrelations 
of the impact parameter selection with the
temperature measurement. At lower bombarding energies
the associated charged particle multiplicity was 
sampled, at forward angles
with a set of 36 CaF$_2$-plastic telescopes of the 
phoswich type, called 
Zero Degree Hodoscope (ZDO) arranged symmetrically around 
the beam axis, and
around 40$^o$ with a set of 6 Si-strip detectors
which, in total, represented 48 channels for 
charged-particle
detection.

\section{Data analysis}
The calibration of more than 200 detectors
was based on measurements with
$\alpha$ and $\gamma$ sources, on the observed 
punch-through points,
and on  the correlation of the responses of the Si and 
CsI detectors
to  particles of known identity and energy.

The coincidence spectra, plotted as a function of the 
relative 
momentum of the two coincident fragments, exhibit peaks 
due to
resonance  decay superimposed on the background of 
coincident but
uncorrelated  particle emission. In order to determine 
the background
spectrum the  technique of correlation functions is 
used. Correlation
functions are  created by dividing the coincidence 
yields by
uncorrelated yields obtained from event-mixing. This 
eliminates the
major part of non-interesting  correlations, e.g. due to 
detector
properties, which permits a more reliable estimate of 
the background.
The background correlation function was determined with 
the
aid of model calculations and by comparing to the 
correlation
functions of other pairs of particles which are free of 
resonances in
the region of small relative momenta, e.g. d + d.

\begin{figure}[htb]
 \centerline{\epsfig{file=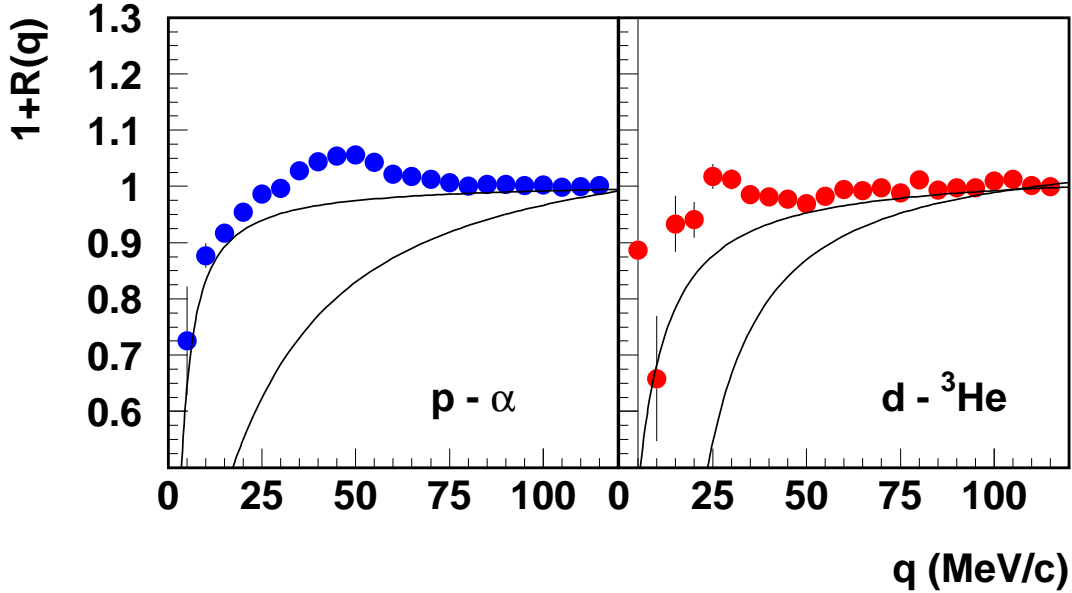,width=\textwidth,
  bbllx=0,bblly=200,bburx=480,bbury=480}}
  \caption[]{
  Correlation functions for the $^5$Li groundstate (left side) and
  the 16.7 MeV excited state (right side) in Au+Au collisions at
  150 AMeV. The lines show the minimum and maximum assumptions to the
  coulomb background.
  }
  \label{corrfunc}
\end{figure}

Typical experimental correlation functions for 
p~-~$\alpha$ and d~-~He$^3$
coincidences, for the reaction Au + Au at 150 MeV per 
nucleon, are
shown in Fig. \ref{corrfunc}. Both correspond to decay channels of 
$^5$Li
fragments, and the peaks due to the ground and excited 
(16.7 MeV)
states are clearly visible. The solid lines represent 
the estimated
background which, at large q, is close to one and, at 
small q,
decreases due to the mutual Coulomb repulsion of the two 
charged
particles.

The acceptance of the hodoscopes for a particular 
resonance decay
depends on the properties of the resonance and on the 
geometry and
resolutions of the detectors. Monte-Carlo simulations 
are best suited
for the determination of the efficiency matrix. 
The modelling
of  additional effects such as multiple scattering in 
the target and
in the  detectors has been incorporated. 

From the acceptance corrected yields the nuclear 
temperatures (emission temperature) can be calculated via:
\[
\frac{Y_1}{Y_2} = \frac{2J_1+1}{2J_2+1} \; e^{-\Delta 
E/T}. 
\]
Here, $Y_i,J_i$ are the yields and the spins of two 
states in the same
isotope, e.g. the two states in $^5$Li, separated by the 
energy
$\Delta E$. This assumes that the states of the decaying 
fragments are
populated according to a Boltzmann law. 

We note that the
experimentally determined  yields may contain 
components due
to the sequential decay of  heavier fragments. Model calculations show, 
however, that the
case of the proton rich nucleus $^5$Li is particularly 
favourable because 
the distortion by sequential feeding is predicted to be 
small\cite{SCH93}.
Clearly, it will be necessary to test  this assumption 
and, more
generally, to verify the assumption of a  statistical 
population by
attempting a consistent description of all observed 
resonance yields
on the basis of sequential feeding models
\cite{NAY92,CHE88,NAY89,ZHU92,SCH94}. At the moment, the 
assumptions
made in these models are quite crude and need some 
improvements.

On the other hand we 
extract temperatures from the double ratios of isotope 
yields,
measured with the telescopes.
Energy spectra of the individually resolved isotopes are 
reproduced by using
a standard three moving sources fit procedure for 1 A GeV 
data, and including a
collective radial flow contribution for the lower energy 
ones.
The ($^3$He/$^4$He)/($^6$Li/$^7$Li) double yield ratios 
are deduced from
the integrated cross sections. We want to remark that 
the flow parameters 
obtained from the fit procedure at the lower energies 
are in good agreement 
with the ones presently available in the 
literature.\cite{LISA,JEO94}

\section{Results and discussion}
\begin{figure}[t]
\begin{minipage}[t]{0.48\linewidth}
\epsfxsize=\linewidth
  \centerline{\epsffile[0 0 360 300]{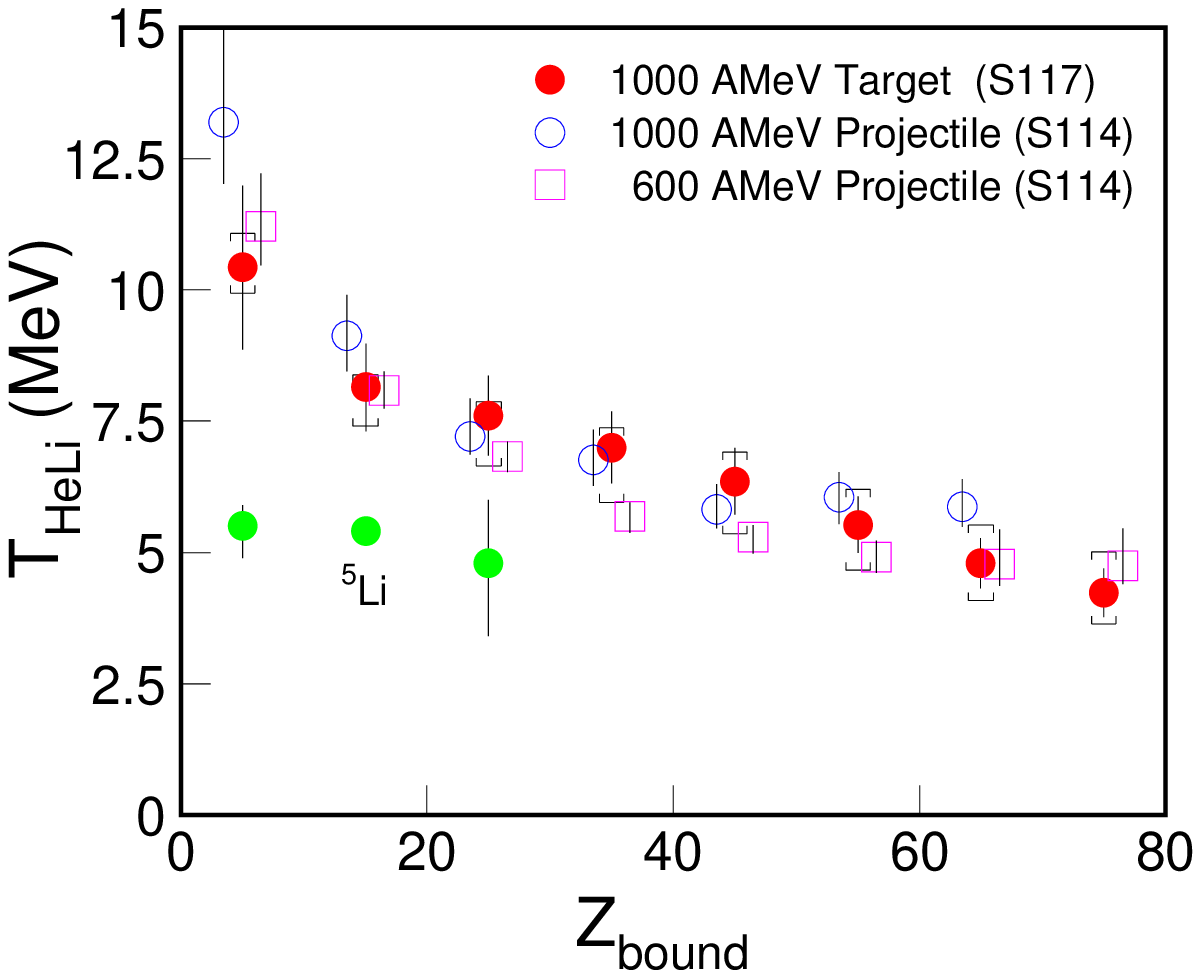}}
  \caption[]{
    Target spectator isotope temperatures T$_{He-Li}$ measured in the
    Au+Au reaction at 1 A GeV are shown (closed symbols) as a function
    of $Z_{bound}$. The open symbols refers to the values obtained for the
projectile spectator in 
     previous experiments.The shadowed symbols refer to the
emission temperatures from $^5$Li excited state population, measured for target
spectator fragmentation at 1 A GeV.
  }
  \label{tperiph}
\end{minipage}
\hfill
\begin{minipage}[t]{0.48\linewidth}
\epsfxsize=\linewidth
\centerline{\epsffile [0 0 360 300]
{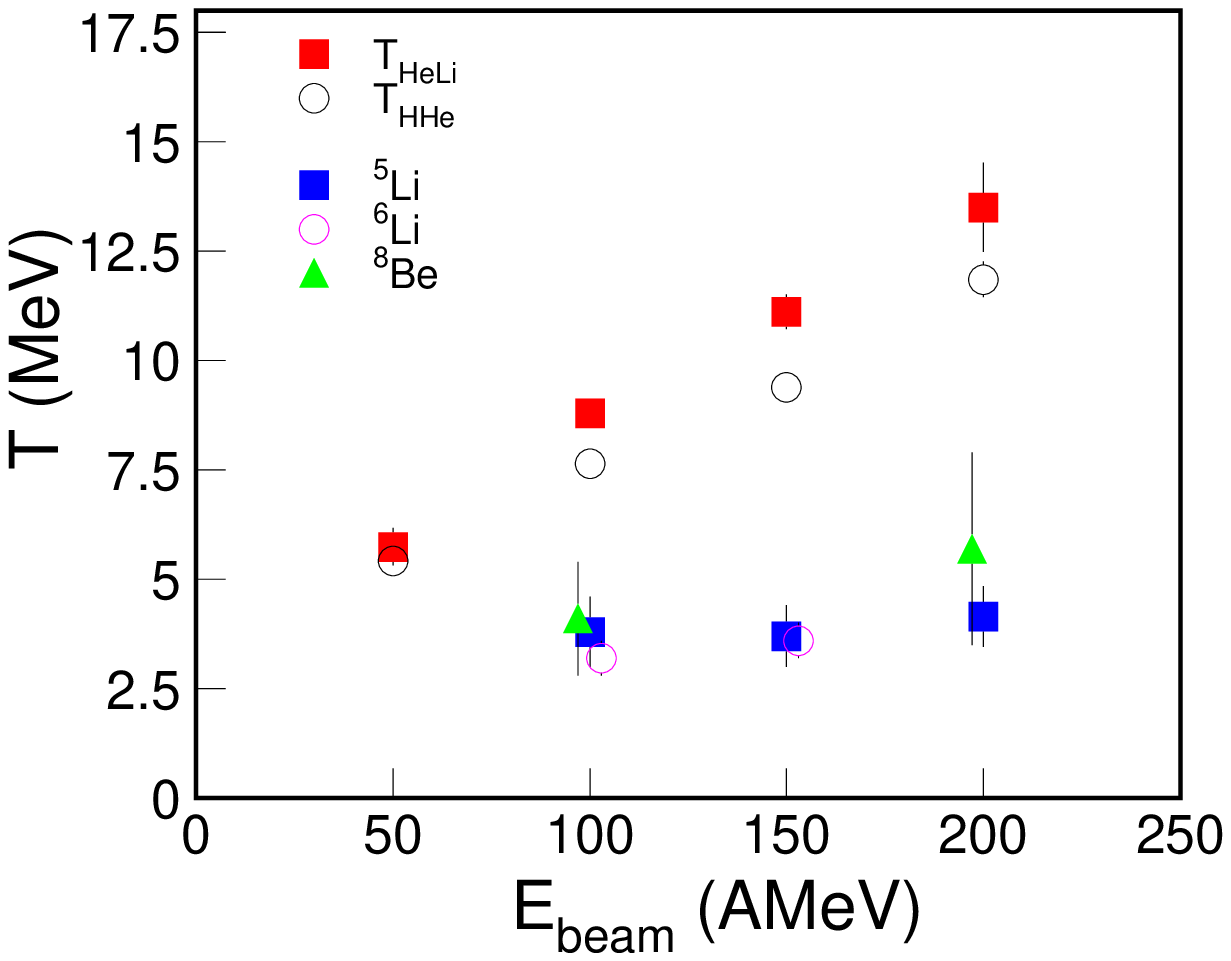}}
  \caption[]{
    Temperatures extracted from the excited states of light nuclei and
    the double ratio of the yields from different isotopes
    are shown as a function
    of the bombarding energy for central Au+Au collisions.The calibration
    of the isotope thermometers was done accordingly to Ref.26.  
  }
  \label{tcentra}

\end{minipage}
\end{figure}

The temperatures extracted from the double ratios of 
($^3$He/$^4$He)/($^6$Li/$^7$Li) isotopes (T$_{He-Li}$ ) emitted from 
the target spectators 
in the reaction Au+Au at 1 A GeV are shown in Fig. \ref{tperiph} as a 
function of  $Z_{bound}$
of the projectile spectators.They are in a very good 
agreement with the values
obtained in a previous experiment\cite{SCHU} for the 
same system but
where isotope temperatures were evaluated for 
projectile spectators. 

Because of the $Z_{bound}$ selection on projectile spectator 
products, the observed 
agreement between the two measurements represents an 
autocorrelation free 
methodical test.The temperatures measured for target 
spectators
vs excitation energies associated with the selected 
$Z_{bound}$ values agree well
with the rising branch of the caloric curve obtained for 
the projectile spectators produced in the Au+ Au 
reaction at 600 A MeV. \cite{POC96}

The values of 
T$_{He-Li}$ and T$_{H-He}$ (($^2$H/$^3$H)/($^3$He/$^4$He)) measured for central 
Au+Au collisions at incident energies of 50,100,150,200 
A MeV are reported in Fig.4 versus the bombarding energy because in the 
present experiment excitation energies are not measured.
The good agreement between the T$_{He-Li}$ and T$_{H-He}$ shows that the 
increasing behaviour of the "isotope" temperature is not very much affected by
the sequential feeding.On the other hand 
 the caloric curves for 
spectator and 
participant nuclear matter are also consistent \cite{POC96}suggesting 
that we are 
dealing with similar
processes.
In order to crosscheck and calibrate the isotope 
temperatures with an 
alternative experimental thermometer, they are compared 
in Fig.3 and 4 with
temperatures deduced from the relative population of 
states in $^5$Li, $^6$Li and $^8$Be.
In peripheral collisions,shadowed circles in
 Fig 3,we notice a slightly increasing discrepancy 
between the two thermometers
as the excitation energy increases (lower $Z_{bound}$ values).
Both thermometers look consistent again if one apply the corrections evidenced
by a systematic study on the effects of the sequential decay feeding 
\cite{BET96}.
In central collisions (Fig.4),without the corrections of Ref.25,
 differences up to 9 MeV between 
the two methods are
reached at the highest bombarding energy.Such a discrepancy is 
independent from the particular choice of the excitation 
energy scale and cannot be accounted for by sequential decay because of
the constancy of the emission temperature
at a very low value of 4 MeV, against an increase of the 
beam energy of a 
factor of four.
At present we have no quantitative explanation for 
the observed differences between the two methods.
It could well be that the
emission temperature method is insentitive to the 
variation of the real
temperature or that the two methods select different 
stages of the decay process.
It is particularly intriguing that the discrepancy between the two thermometers 
arises at beam energies where collective radial 
flow become important. Whether this is a pure coincidence or whether
there exists some causality remains to be seen.
We conclude that while the recent ALADIN experiment confirms the 
earlier results on the isotope temperatures, the different 
energy dependence of the isotope and emission temperatures seems 
to exclude a common freeze-out point for chemical and internal 
degrees-of-freedom.


\section*{Acknowledgments}
This work was supported in part by the European 
Community under contract ERBCHGE-CT92-0003 and ERBCIPD-CT94-009. J.P. and M.B.
 acknowledge the financial support of the DFS under the contract No. Po256/2-1 
and No. Be1634/1-1 respectively.

\end{document}